\documentclass[12pt,preprint]{aastex}

\shorttitle{GRB Pulse Properties}
\shortauthors{Hakkila et al.}

\begin{document}

\title{Correlations Between Lag, Luminosity, and Duration in Gamma-ray Burst Pulses}

\author{Jon Hakkila\altaffilmark{1}, Timothy W. Giblin\altaffilmark{2}, Jay P. Norris\altaffilmark{3}, \\
P. Chris Fragile\altaffilmark{1}, and Jerry T. Bonnell\altaffilmark{4}}
\affil{$^1$Dept. Physics and Astronomy, The College of Charleston, Charleston, SC  29424-0001\\$^2$Dept. Physics and Astronomy, The University of North Carolina, Asheville, NC \\$^3$Space Science Division, NASA/Ames Research Center, Moffett Field, CA\\$^4$UMCP/CRESST/GSFC - Greenbelt, MD}
\email{hakkilaj@cofc.edu}

\begin{abstract}

We derive a new peak lag vs. peak luminosity relation in gamma-ray 
burst (GRB) pulses. We demonstrate conclusively that GRB spectral 
lags are {\em pulse} rather than {\em burst} properties and show how the 
lag vs.\ luminosity relation determined from CCF measurements of 
burst properties is essentially just a rough measure of this newly 
derived relation for individual pulses.  We further show that most 
GRB pulses have correlated properties: short-lag pulses have 
shorter durations, are more luminous, and are harder within a burst 
than long-lag pulses. We also uncover a new pulse duration vs. pulse 
peak luminosity relation, and indicate that long-lag pulses often 
precede short-lag pulses. Although most pulse behaviors are 
supportive of internal shocks (including long-lag pulses), we identify 
some pulse shapes that could result from external shocks.

\end{abstract}




\section{Introduction}

Gamma-ray burst (GRB) prompt emission has remained enigmatic 
over the years, even as our understanding of afterglow physics has evolved.
Time intervals containing flux increases ({\em pulses}) exhibit some
general behaviors, including 
(1) longer decay than rise rates, (2) hard-to-soft spectral evolution, and 
(3) broadening at lower energies (e.g. \citet{nor96, ryd05a}).

The {\em spectral lag} is the delay between photons observed in a high-energy
bandpass relative to a lower-energy one; it is primarily obtained through
application of the cross-correlation function (CCF; \citet{ban97}).
In general, lag is an indicator of both GRB peak luminosity 
(e.g. \citet{nor00, nor02}) and time history morphology \citep{hak07}, 
with short-lag, variable bursts having greater luminosities than 
long-lag, smooth bursts.

The short durations, spectral evolution, and short interpulse durations
of most GRB pulses suggest that they originate from
{\em internal} shocks in relativistic winds 
(e.g. \citet{dai98, rrm00, nak02}). However, some bursts exhibit a 
soft component indicative of afterglow onset (e.g. \citet{gib02})
which could be interpreted as the initial {\em external} shock. 
Prompt afterglow emission begins preferentially towards the end of the 
burst or even after the GRB ends, as suggested by 
several x-ray afterglows observed by BeppoSAX \citep{cos00} and Swift 
\citep{pan07}, but it can also appear at a time early enough to 
overlap the short-timescale emission (as observed in GRB 980923 
\citep{gib99}). In addition, co-adding fluxes of many BATSE GRBs has
led to observation of extended soft gamma-ray emission possibly indicating
the same phenomenon \citep{con02}.

Internal and external shocks have both been predicted theoretically (e.g. \citet{sar99}), 
and it has been suggested that both shock types are observed within 
quiescent BATSE GRBs 960530 and 980125 \citep{hak04}. 
{\em Quiescent} GRBs release their prompt emission in more than one distinct 
{\em emission episode} (each episode contains one to many overlapping pulses).
Although the pre-quiescent emission episodes of these two GRBs
satisfy the standard internal 
shock paradigm, the long episodic lags, smooth morphologies, and soft spectral
evolution of the later episodes are more consistent with external shocks, 
indirectly implying that the pulses comprising these episodes might also 
relate to specific types of shocks (however, see \citet{chi07}). 
Other observations support the idea that lag can vary within GRBs,
and indirectly suggest that these variations are associated with pulses
(e.g. \citet{nor02, ryd05b, che05}). However, no study has yet
isolated and delineated pulse spectral properties; to this end and
to test the hypothesis that short-and long-lag pulses have different origins, 
we have set out to model pulses and study their spectral dependences.

\section{Pulse Identification and Fitting Technique}

We have developed a semi-automated pulse-identification
and fitting routine using BATSE 64 ms data. The Bayesian Blocks 
(BB) routine \citep{sca98} is applied to summed 4-channel data 
and identifies regions over which counts change significantly. 
Each BB potentially contains a pulse, and is fit using the pulse 
model of \citet{nor05}: 
\begin{equation}
I(t) = A \lambda \exp^{[-\tau_1/(t - t_s) - (t - t_s)/\tau_2]}
\end{equation}
where $t$ is time since trigger, $A$ is the pulse amplitude, 
$t_s$ is the pulse start time, $\tau_1$ 
and $\tau_2$ are characteristics of the pulse rise and pulse decay, and 
$\lambda = \exp{[2 (\tau_1/\tau_2)^{1/2}]}$. 
Additionally, a two-parameter linear background model is assumed.
Reasonable initial guesses are made using the starting and ending 
BB boundaries and the highest flux found in this BB. The nonlinear least 
squares routine MPFIT ({\tt http://idlastro.gsfc.nasa.gov/contents.html}) 
simultaneously fits all of the pulses and the background, iterating 
initial guesses to a convergent solution as characterized by
$\chi^2$ per degree of freedom. Statistically insignificant pulses are 
deselected from the initial pulse fit using a dual-timescale threshold 
\citep{hak03} and Occam's Razor. The dual-timescale threshold is chosen 
over a peak flux (favoring short, intense pulses) or fluence 
threshold (favoring long, low-intensity pulses) because it treats equally 
pulses peaking on short and long timescales.  The reduced model is run 
again with fewer pulses until it converges and only pulses brighter than 
the dual-timescale threshold remain.
The 4-channel pulse characteristics are used as starting points from
which individual energy channel pulse characteristics are obtained. 
The process described above is repeated until convergent solutions 
are obtained for pulses in each energy channel.

{\em Pulse peak lags} are defined as the differences between the pulse 
peak times in different energy channels (pulse peak times are given by
$\tau_{\rm peak} = t_s + \sqrt{\tau_1 \tau_2}$).
Pulse peak lags can be obtained for any pulse between two energy 
channels, although we define the standard $l_{pp}$ as that measured 
between energies of 100 to 300 keV (BATSE channel 3) and 25 to 50 keV 
(BATSE channel 1). Other measurable pulse properties include
the pulse width $w = [9 + 12\sqrt{\tau_1/\tau_2}]^{1/2}$
and the pulse asymmetry $\kappa = w/(3 + 2\sqrt{\tau_1/\tau_2}).$
These definitions are based on time intervals 
between intensities of $A {e^{-3}}$, rather than of $A {e^{-1}}$ \citep{nor05}. 
We note that the {\em modeled} pulse duration definition of $w$ is less 
susceptible to statistical variations near the beginning and end of the pulse 
than observationally-defined durations such as $T_{90}$.

Pulse parameter uncertainties are obtained using Monte Carlo analysis because
Gaussian error assumptions \citep{nor05} is not always valid for $t_s$, $\tau_1$, 
and $\tau_2$. These distributions often resemble lognormal or multi-modal 
distributions, and improperly-quoted uncertainties in these parameters
often lead to overestimated uncertainties for pulse peak time amplitudes, 
durations, and asymmetries.

\section{Analysis}

Energy-dependent pulse properties are regularly identified and extracted
using this technique,
even though the pulse-finding technique is energy-independent. 
Multi-lag GRBs are apparently the norm for most Long GRBs.
When the signal is strong enough that pulses can be cleanly and unambiguously 
extracted, their characteristics are generally consistent across all energy 
channels (e.g.\ lags are observed across all energy channels in proportion to the 
energy channel separation).  Within the limit of uncertainty, it appears that 
every pulse is characterized by its own lag. This is not to say that all pulses 
are clearly identified from one energy channel to another; there are fitting 
ambiguities caused by pulse evolution, pulse overlap,
and low signal-to-noise (such as is found in BATSE channel 4). 
Additionally, it is difficult to uniquely fit many overlapping pulses 
because a pulse's fitted signal-to-noise is not solely dependent on background; 
the flux of other pulses is ``noise" to this fit. 

\begin{figure}
\plottwo{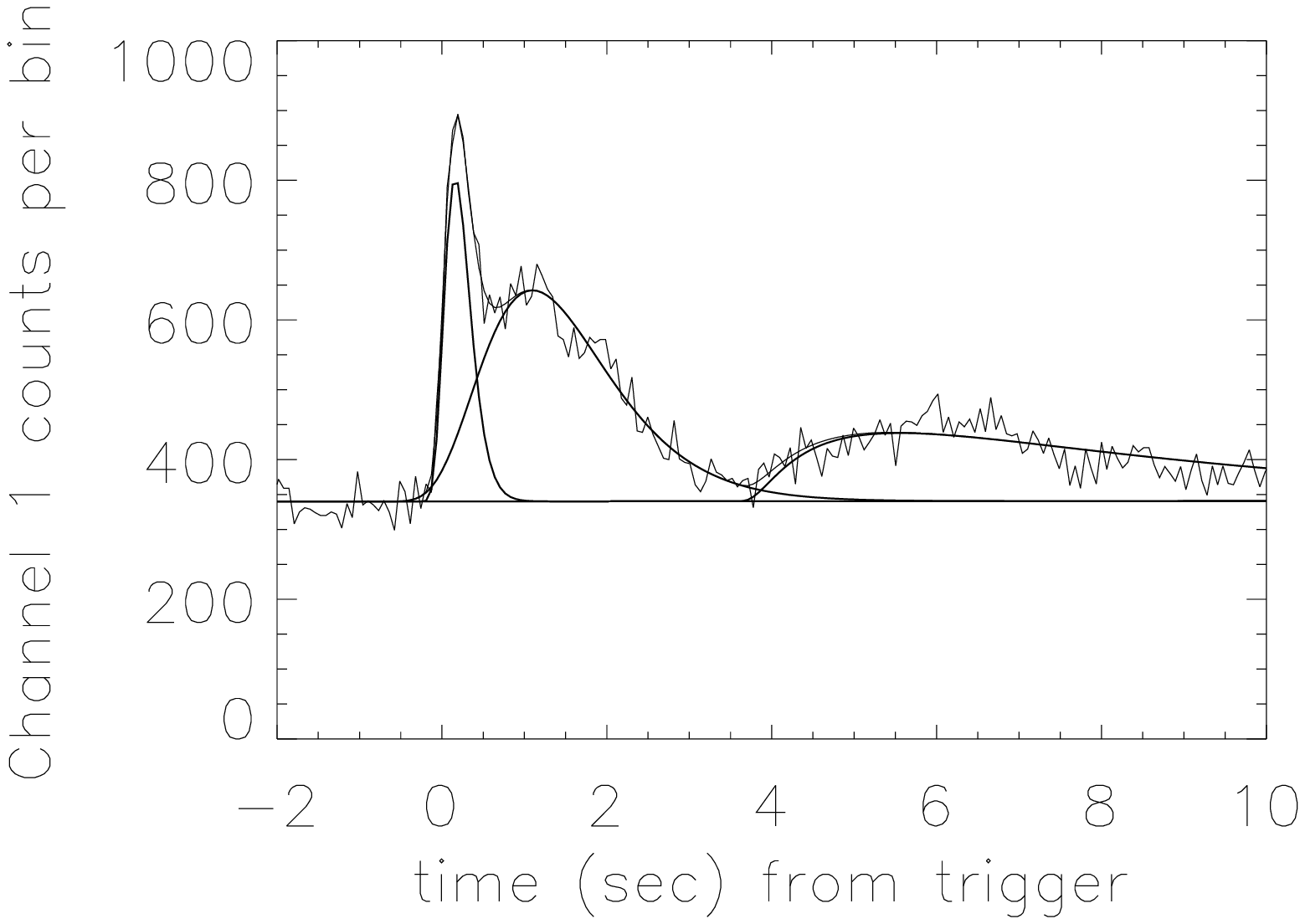}{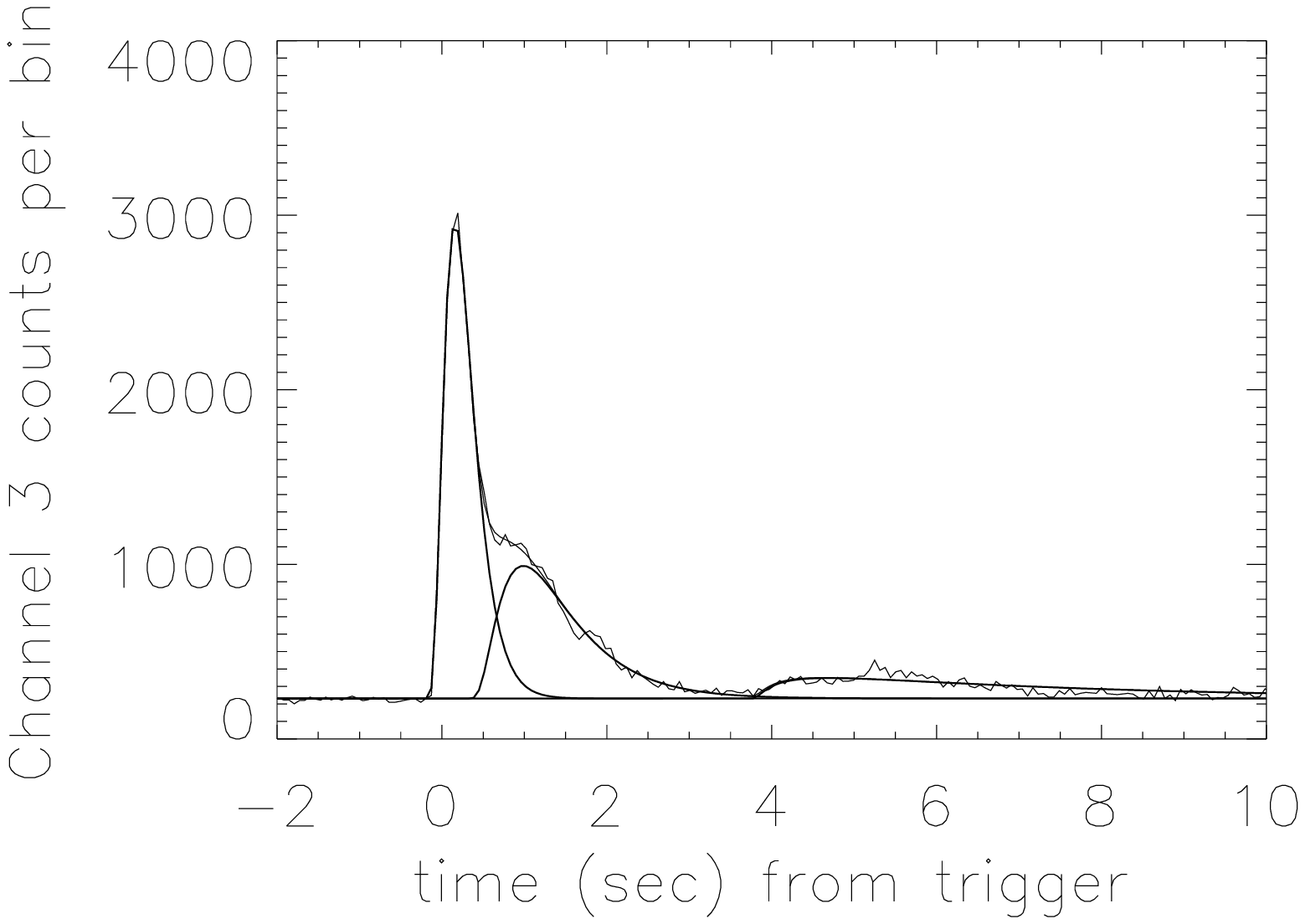}
\caption{GRB 950325a pulse fits for BATSE channel 1 (first panel) and 
channel 3 (second panel). \label{fig1}}
\end{figure}

We demonstrate results with the fit to GRB 950325a (BATSE trigger 3480),
in which three overlapping pulses are 
observed with fairly high signal-to-noise in all four BATSE energy channels.
Fitted channel 1 and channel 3 time histories are shown in Figure \ref{fig1}.  
The CCF burst lag $l_{pp} = 0.014 s \pm 0.009$ \citep{hak07}.
Some properties of the extracted pulses are listed in Table \ref{tbl-1};
these include the pulse fit parameters $l_{pp}$, $w$, and $\kappa$
as well as CCF pulse lags. 
Pulse CCFs have been reconstructed from multichannel 
pulse parameters and the burst's Poisson background. 
The greatest contribution to the CCF lag 
is found to come from the shortest-lag, highest-intensity pulses. 
In fact, the CCF lag appears to be insensitive to the presence of longer-lag pulses,
which can appear at any time during a burst. The CCF and pulse peak lags
occasionally disagree: the CCF lag of pulse 2 is demonstrably negative, 
even though the pulse has a moderately long pulse lag that is measured 
consistently across all four energy channels. We are currently exploring 
the effects of pulse shape on the CCF. The CCF of the original burst is 
essentially identical to that of the reconstructed 
burst, indicating that the long- and short-lag pulses have been 
combined to reproduce a good facsimile of the original short-lag GRB.


The sensitivity of the CCF to the shortest, brightest pulses provides an
explanation for why \citet{nor00} finds different CCF lags when their burst
data are confined to different intensity regimes via an apodization technique. 
Their removal of the low-intensity flux predominantly 
removes CCF contributions from the longest duration, longest-lag pulses,
and improves the ability to measure the signal from the short-lag pulses. 

\subsection{Spectral-Dependent Pulse Properties}

Although we are in the early stages of analyzing GRB
pulse data, it is already apparent that the vast majority of GRB pulses 
have correlated spectral and temporal properties. 
Pulse lag, amplitude, duration, and hardness
are linked; correlations between these behaviors lead 
us to believe that most GRB pulses represent a single physical 
phenomenon. Pulses exhibiting possible exceptions to these behaviors 
are those that cannot be appropriately fit using these techniques. 
These pulses tend to fall into three categories: (1) pulses in crowded 
fields that cannot be unambiguously resolved (many have 
very short durations), (2) low signal-to-noise pulses that cannot be 
unambiguously resolved, and (3) bright yet uncommon pulses having shapes 
not adequately fitted by the four-parameter pulse model; when we are able
to force a fit, we find these pulses to typically have short rise and very 
long decay times.
We cannot
ascertain if there are systematic biases in pulse sampling,
because we cannot know the properties of pulses we cannot measure. 
However, we note that we preferentially cannot fit pulses in complex GRBs, 
as these bursts by definition have many overlapping pulses.

Two key correlations are identified in a sample of 24 pulses from
13 BATSE bursts \citep{hak08}: {\em (1) Pulse amplitude} (measured across BATSE's 
four-channel energy range in units of counts $\rm sec^{-1}$) {\em decreases 
with increasing lag} (a Spearman rank-order correlation test indicates 
a probability of only $2.4 \times 10^{-8}$ that this correlation could 
occur randomly), and {\em (2) Pulse width $w$} (measured across 
BATSE's four-channel energy range) {\em increases with increasing lag} 
(the Spearman rank-order probability is $1.7 \times 10^{-7}$).
It is remarkable that these correlations are robust enough to be clearly 
identified in the observer's rather than in the GRB rest frame; intrinsic 
pulse properties (e.g. jet properties) must be significantly larger than 
extrinsic effects (e.g. cosmological redshift and the inverse square law) for 
this to be true. Short-lag pulses are brighter and shorter than long-lag pulses; 
this appears to be true both within bursts and from one burst to another.

\begin{figure}
\plotone{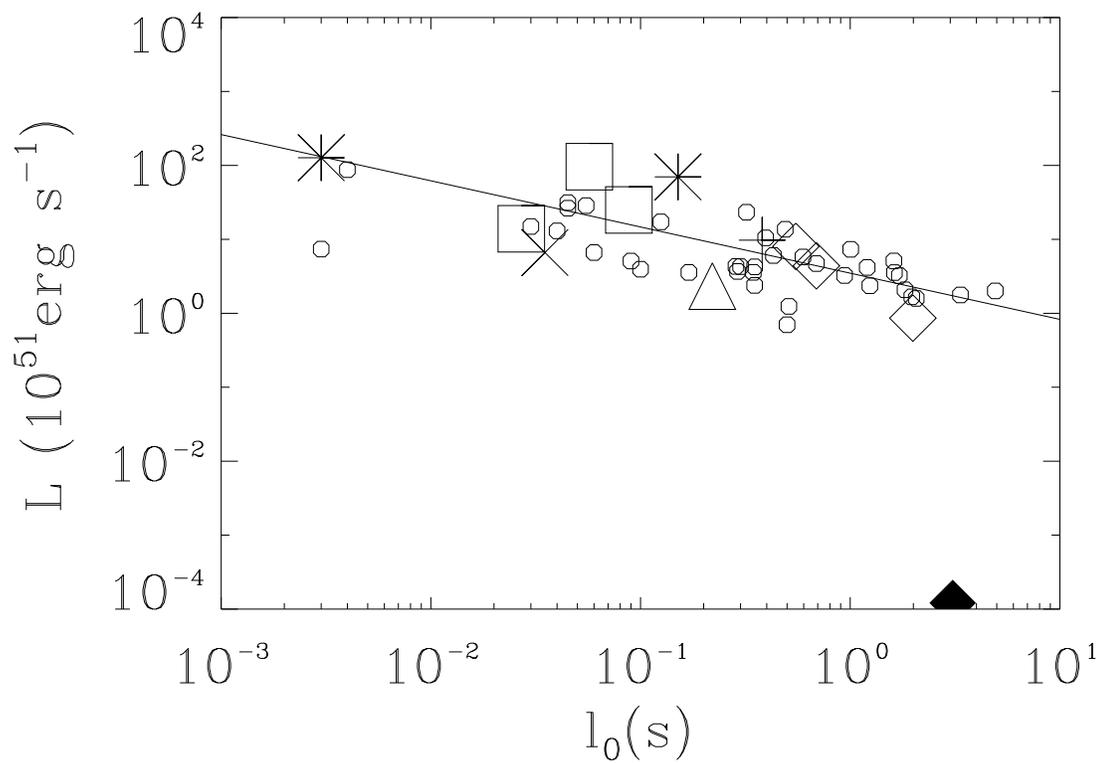}
\caption{Isotropic pulse peak luminosity $L$ vs.\ pulse peak lag 
($l_{0}$) for fit pulses of BATSE GRBs having known luminosities. 
The sample consists of pulses from GRB 971214 (asterisk), GRB 980703 
(open diamond), GRB 970508 (triangle), GRB 990510 (square), GRB 
991216 (X), and GRB 990123(plus), and the underluminous GRB 980425 
(filled diamond). Symbol size denotes approximate uncertainty. Also plotted
are 38 pulses from 22 BATSE GRBs without known redshifts (small circles); 
$z=1$ is assumed. \label{fig2}}
\end{figure}

\begin{figure}
\plotone{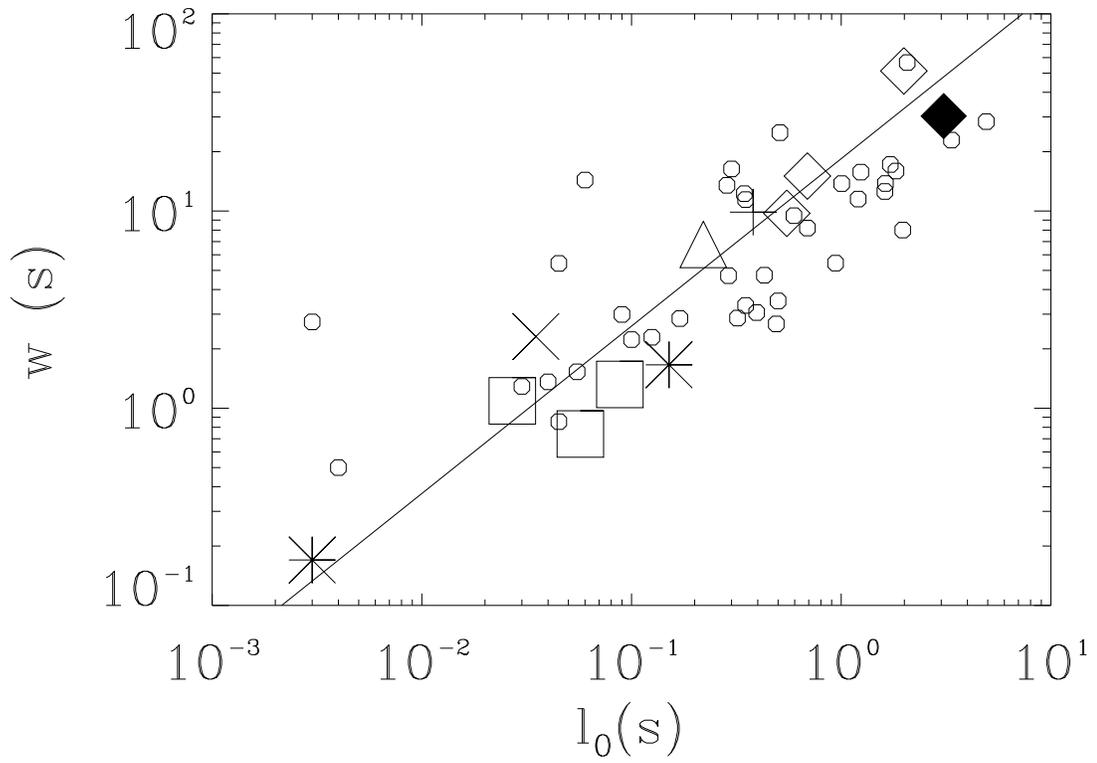}
\caption{Pulse duration $w$ vs.\ pulse peak lag ($l_{0}$) for pulses shown 
in Figure \ref{fig2}.\label{fig3}}
\end{figure}

\begin{figure}
\plotone{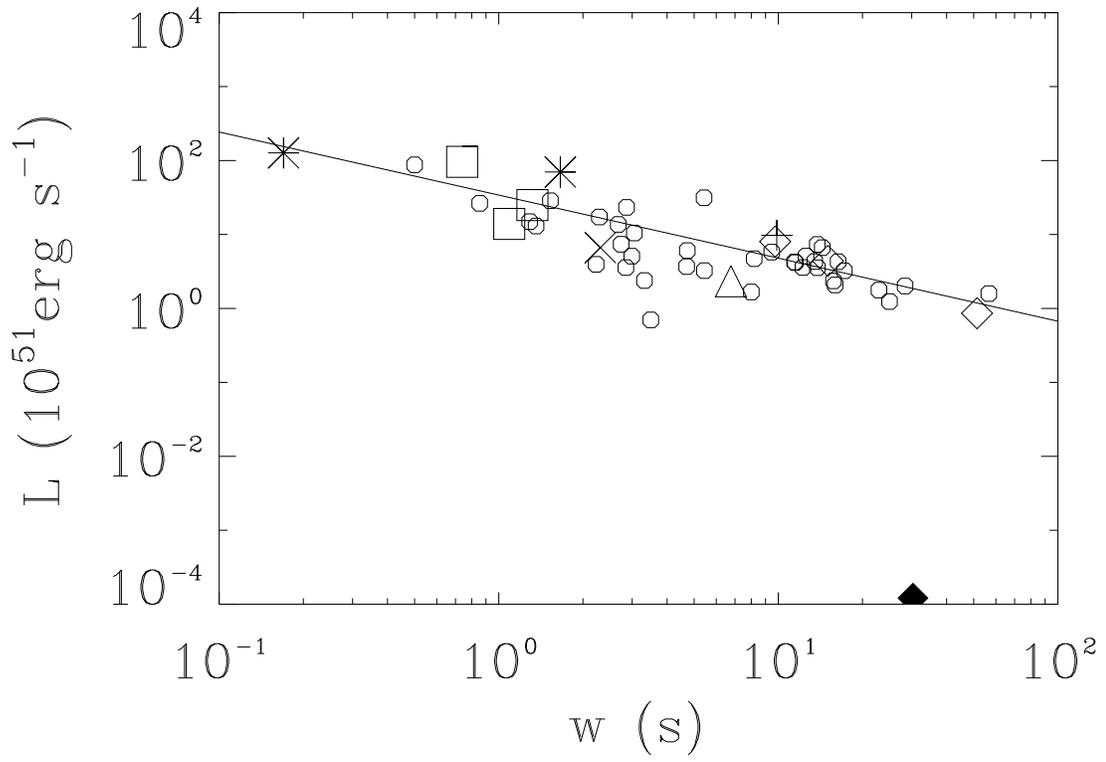}
\caption{Pulse duration $w$ vs.\ isotropic pulse peak luminosity $L$ for pulses shown 
in Figures \ref{fig2} and \ref{fig3}.\label{fig4}}
\end{figure}

To clarify what these relationships mean, we fit pulses of BATSE bursts 
with known redshift; these GRBs originally defined the lag vs.\ peak luminosity 
relation \citep{nor00}. Some pulses in these GRBs cannot 
be fitted due to poor resolution and/or pulse overlap, so several complex GRBs 
have been limited to only one or two isolated pulses each. The resulting sample
consists of 12 pulses in 7 bursts. Fitted peak fluxes have been re-calibrated on the 
256 ms time scale so that the results can be compared to those of \citet{nor00}. 
The pulse peak lag, pulse duration, and pulse 
intensity have been corrected to the GRB rest frame. The pulse characteristics 
are plotted in Figures \ref{fig2} through \ref{fig4}. Also plotted
are pulses from BATSE GRBs without known redshifts, assuming $z=1$. 
These pulses demonstrate 
1) the pervasiveness of the correlations, and 2) the strength of these
intrinsic relationships relative to cosmological effects. 

Figure \ref{fig2} demonstrates the pulse peak luminosity $L_{51}$ (the 
isotropic pulse peak luminosity $L$ in units of $10^{51}$ erg s$^{-1}$) 
vs. the rest frame pulse peak lag $l_0$ (obtained by shifting $l_{pp}$ 
into the rest frame); this is similar to the \citet{nor00} lag vs.\ peak luminosity 
diagram, except that it has been applied to {\em pulses} rather than to the 
bursts themselves.  The best-fit functional form of  this relation (excluding
underluminous GRB 980425) is $\log(L_{51}) = A + B \log(l_0)$ 
($A$ is in units of $10^{51}$ ergs s$^{-1}$). An anti-correlated relationship 
(correlation coefficient $R=-0.72$)
is identified, with $A= 0.54 \pm 0.05$ and $B= -0.62 \pm 0.04$.
The validity of this relation for pulses both within and across GRBs
implies that the pulse peak lag vs.\ pulse peak luminosity relation is a 
fundamental one,
while the CCF lag vs.\ peak luminosity relation is of secondary importance.
In fact, the two relationships have different power-law indices, with
the lag vs.\ peak luminosity relation's power-law index being 
$B=-1.14 \pm 0.10$ \citep{nor00}.  We have already demonstrated that 
the CCF lag is merely a rough measure of the narrowest pulse's lag. 
Similarly, the peak flux is an overestimate of the narrowest pulse's amplitude: 
peak flux results from summing fluxes from overlapping 
pulses, Poisson errors also make the measured peak fluxes
larger than the modeled fluxes (the ``Meegan Bias").


Figure \ref{fig3} plots the rest frame pulse duration ($w_0$; given in units of s) 
vs.\ pulse lag ($l_0$). Remarkably, the pulse duration vs.\ pulse lag correlation holds 
over four orders of magnitude in both duration and lag; even the underluminous 
GRB 980425 follows it. Its functional form in the GRB rest frame is taken as
$\log(w_0) = C + D \log(l_0)$. From this small sample, the values of the 
coefficients are $C= 1.27 \pm 0.01 $ s and $D=0.85 \pm 0.01$, with $R=0.95$.


The strong correlations in Figures \ref{fig2} and \ref{fig3} indicate that there 
must also be a {\em pulse width vs.\ pulse peak luminosity} relation. Figure 
\ref{fig4} demonstrates this relation, which is more tightly defined ($R=-0.88$)
than the pulse peak lag vs.\ pulse peak luminosity relation. Again, the exception 
to the rule is underluminous GRB 980425. The best-fit functional form of 
this relation in the GRB rest frame is $\log(L_{51}) = E + F\log(w_0)$, with
$E= 1.53 \pm 0.02$ and $F=-0.85 \pm 0.02$ ($E$ in units of $10^{51}$ 
ergs s$^{-1}$).


For GRBs with multiple fitted pulses, we also find that pulse spectral hardness
(channel 3 pulse fluence divided by channel 1 pulse fluence)
anti-correlates with pulse lag and duration, and correlates 
with pulse intensity. The same correlations are found in GRB 950325a,
and imply that spectral evolution is present both across pulses and within them.

These results also allow new insights into the low peak luminosity of GRB 
980425. This GRB's single pulse is similar in lag and duration to the 
long-lag, long duration pulse of GRB 980703, implying a similar physical 
mechanism. Yet, GRB 980425 is still four orders of magnitude less luminous,
similar to the underluminous, long-lag, long duration XRF 060218 \citep{lia06}.
Although \citet{geh06} postulate a separate class of underluminous
GRBs, pulse properties indicate that the low luminosity could still be due to a
purely observational effect, such as large viewing angle relative to the jet \citep{sal01}.

The apparently universal nature of most GRB pulses also explains why the 
CCF lag and the internal luminosity function power-law index (ILF) are 
excellent GRB time history morphology indicators \citep{hak07}. 
CCF lags indicate the presence of short-lag pulses, while the ILF is a 
sensitive indicator of the number of pulses and of pulse shape.


The short durations of most GRB pulses, along with the similar behaviors
seen in these pulses, argue that most pulses are related to internal shocks
rather than external shocks. Since most pulses 
do not seem to have external shock characteristics, we look elsewhere to find
these pulses; e.g.\ to pulses that cannot be fitted easily using the pulse model. 
Although we cannot say anything about faint or overlapping pulses, 
we suggest that pulses with short rise and very long decay times might
represent external shocks capable of initiating afterglow. Such pulses appear
to have short lags, although they are often fitted by two overlapping pulses;
one short, large-amplitude, short-lag pulse and one long, 
small-amplitude, indeterminate-lag pulse. Suggestive of these pulse
types is the extended tail in GRB 980923 \citep{gib99}, thought to occur at the 
transition from prompt emission to afterglow. We are exploring this hypothesis.

\section{Conclusions}

Multi-lag GRBs are ubiquitous -- this paper provides the first clear evidence
in support of this statement. Each pulse appears to be
characterized by its own lag; lag is a consequence of pulse evolution 
rather than a burst property. Burst peak luminosity and the CCF lag 
are not fundamental properties, but result from pulse combinations. 
Pulses are the basic, central building blocks of GRB prompt emission,
and it is essential to our understanding of GRB physics
that we properly catalog and characterize pulse properties. 

Pulse lag, pulse luminosity, and pulse duration strongly correlate, implying that most 
GRB pulses have similar physical mechanisms; these are more
consistent with internal than external shocks. Short pulses 
presumably indicate a collision of material at larger relative Lorentz 
factor than long pulses, and a large Lorentz factor requires a
cleaner fireball with less baryonic matter. The fireball opacity dictates 
the emission timescale, so a clean, 
high-amplitude fireball should have a short decay and a short lag, while
a dirty, low-amplitude fireball should produce a long decay and a long lag.

\acknowledgments

We are grateful to Rob Preece, Tom Loredo, and Jeff Wragg for helpful discussions.  
The material presented here is based upon work supported by NASA under award No. GRNASNNX06AB43G and through the South Carolina NASA Space Grant program.

\clearpage

\clearpage

\clearpage

\clearpage

\clearpage

\begin{table}
\begin{center}
\caption{Pulse Properties for GRB 950325a.\label{tbl-1}}
\begin{tabular}{cccccc}
\tableline\tableline
Pulse & peak (cts s$^{-1})$& $w (s)$ & $\kappa $  & $l_{pp} (s) $ & $l_{\rm CCF} (s)$ \\
\tableline
1 & $5112 \pm 54$ & $1.00 \pm 0.08$ & $0.76 \pm 0.01$ & $0.004 \pm 0.009$ & $0.006 \pm 0.003$\\
2 & $1669 \pm 36$ & $3.06 \pm 0.10$ & $0.87 \pm 0.01$ & $0.111 \pm 0.045$ & $-0.064 \pm 0.020$\\
3 & $  354 \pm   8 $ & $9.45 \pm 0.52$ & $0.93 \pm 0.02$ & $0.854 \pm 0.150$ & $0.414 \pm 0.062$\\
\tableline
\end{tabular}
\end{center}
\end{table}

\end{document}